\documentclass[12pt]{article}
\usepackage[dvips]{graphicx}
\usepackage{epsfig,amsmath}

\begin{document}
\thispagestyle{empty}
\begin{center} 
{\Large\bf Statistical approach of parton distributions:\\ a closer look at the high-$x$ region \footnote{Invited talk at the 3rd Int. Workshop on Nucleon Structure at Large Bjorken $x$, "`HiX2010"', Jefferson Lab., Newport News, Virginia, USA, October 13 - 15, 2010, to appear in the AIP Conference Proceedings}}\\

\vskip 1.4cm
{\bf Jacques Soffer}
\vskip 0.3cm
{\it Physics Department, Temple University},\\ 
{\it Philadelphia, PA 19122-6082, USA}\\
\end{center}
\vskip 1.5cm
\begin{center}
{\bf Abstract}
\end{center}

We recall the physical features of the parton distributions in the quantum statistical approach of the nucleon, which allows to
describe simultaneously, unpolarized and polarized Deep Inelastic Scattering data. Some
 predictions from a next-to-leading order QCD analysis are compared to recent experimental results and we stress the importance of some 
 tests in the high-$x$ region, to confirm the validity of this approach.\\


\vskip 1.4cm
Let us first recall some of the basic ingredients for building up the parton distribution functions (PDF) in the statistical approach, as oppose to the standard polynomial type
parametrizations, based on Regge theory at low $x$ and counting rules at large $x$.
The fermion distributions are given by the sum of two terms \cite{bbs1}, the first one, 
a quasi Fermi-Dirac function and the second one, a flavor and helicity independent diffractive
contribution equal for light quarks. So we have, at the input energy scale $Q_0^2=4 \mbox{GeV}^2$,
\begin{equation}
xq^h(x,Q^2_0)=
\frac{AX^h_{0q}x^b}{\exp [(x-X^h_{0q})/\bar{x}]+1}+
\frac{\tilde{A}x^{\tilde{b}}}{\exp(x/\bar{x})+1}~,
\label{eq1}
\end{equation}
\begin{equation}
x\bar{q}^h(x,Q^2_0)=
\frac{{\bar A}(X^{-h}_{0q})^{-1}x^{2b}}{\exp [(x+X^{-h}_{0q})/\bar{x}]+1}+
\frac{\tilde{A}x^{\tilde{b}}}{\exp(x/\bar{x})+1}~.
\label{eq2}
\end{equation}
Notice the change of sign of the potentials
and helicity for the antiquarks.
The parameter $\bar{x}$ plays the role of a {\it universal temperature}
and $X^{\pm}_{0q}$ are the two {\it thermodynamical potentials} of the quark
$q$, with helicity $h=\pm$. It is important to remark that the diffractive contribution 
occurs only in the unpolarized distributions $q(x)= q_{+}(x)+q_{-}(x)$ and it is absent in the valence $q_v(x)= q(x) - \bar {q}(x)$ and in the helicity
distributions $\Delta q(x) = q_{+}(x)-q_{-}(x)$ (similarly for antiquarks).
The {\it eight} free parameters\footnote{$A=1.74938$ and $\bar{A}~=1.90801$ are
fixed by the following normalization conditions $u-\bar{u}=2$, $d-\bar{d}=1$.}
in Eqs.~(\ref{eq1},\ref{eq2}) were
determined at the input scale from the comparison with a selected set of
very precise unpolarized and polarized Deep Inelastic Scattering (DIS) data \cite{bbs1}. They have the
following values
\begin{equation}
\bar{x}=0.09907,~ b=0.40962,~\tilde{b}=-0.25347,~\tilde{A}=0.08318,
\label{eq3}
\end{equation}
\begin{equation}
X^+_{0u}=0.46128,~X^-_{0u}=0.29766,~X^-_{0d}=0.30174,~X^+_{0d}=0.22775~.
\label{eq4}
\end{equation}
For the gluons we consider the black-body inspired expression
\begin{equation}
xG(x,Q^2_0)=
\frac{A_Gx^{b_G}}{\exp(x/\bar{x})-1}~,
\label{eq5}
\end{equation}
a quasi Bose-Einstein function, with $b_G=0.90$, the only free parameter
\footnote{In Ref.~\cite{bbs1} we were assuming that, for very small $x$,
$xG(x,Q^2_0)$ has the same behavior as $x\bar q(x,Q^2_0)$, so we took $b_G = 1
+ \tilde b$. However this choice leads to a too much rapid rise of the gluon
distribution, compared to its recent  determination from HERA data, which
requires $b_G=0.90$.}, since $A_G=20.53$ is determined by the momentum sum
rule.
 We also assume that, at the input energy scale, the polarized gluon 
distribution vanishes, so $x\Delta G(x,Q^2_0)=0$. For the strange quark distributions, the simple choice made in Ref. \cite{bbs1}
was greatly improved in Ref. \cite{bbs2}. More recently, new tests against experimental (unpolarized and
polarized) data turned out to be very satisfactory, in particular in hadronic
collisions, as reported in Refs.~\cite{bbs3,bbs4}.\\
\begin{figure}[htb]
  \begin{minipage}{7.0cm}
  \epsfig{figure=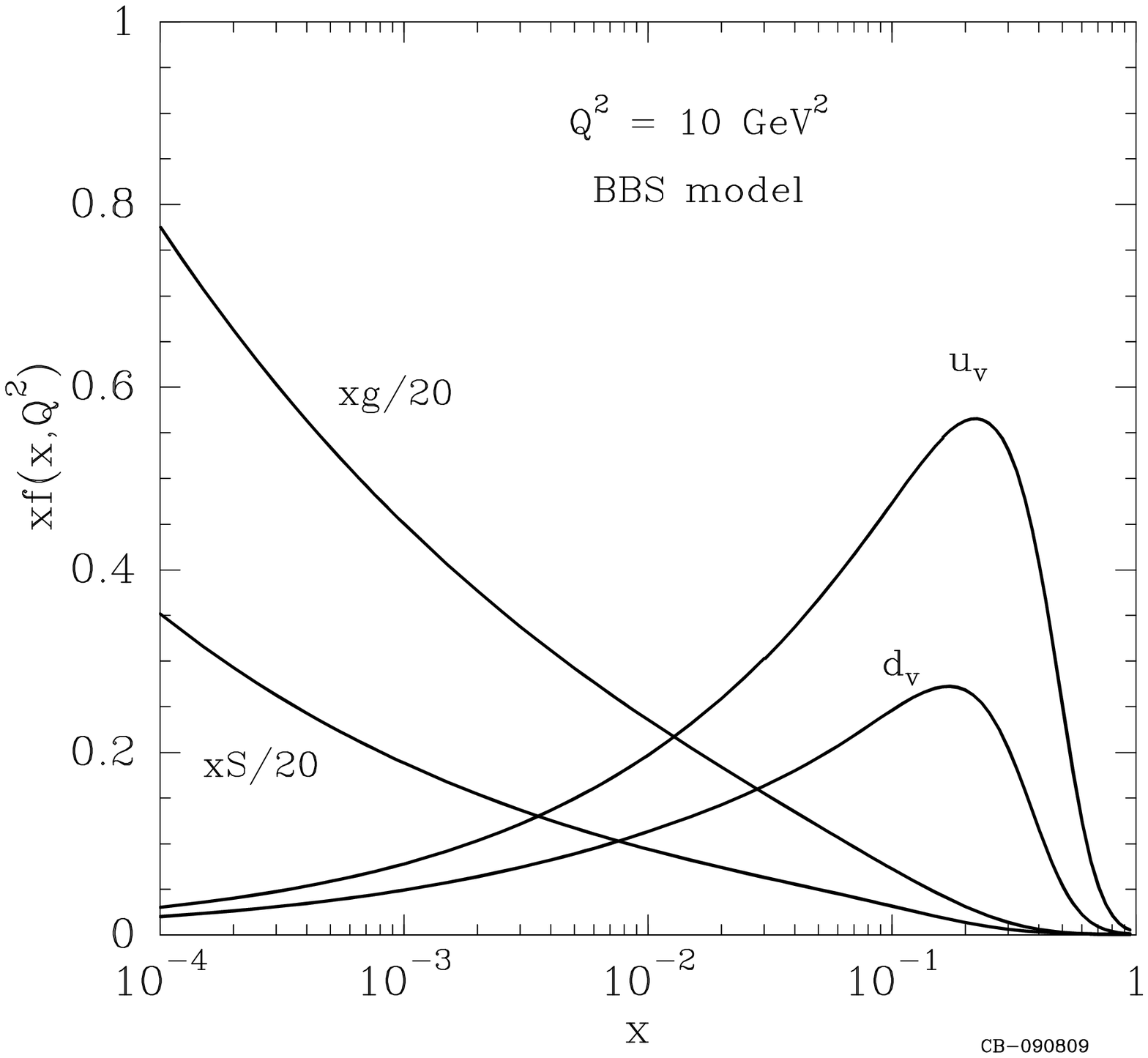,width=7.4cm}  
  \end{minipage}
    \begin{minipage}{7.0cm}
  \epsfig{figure=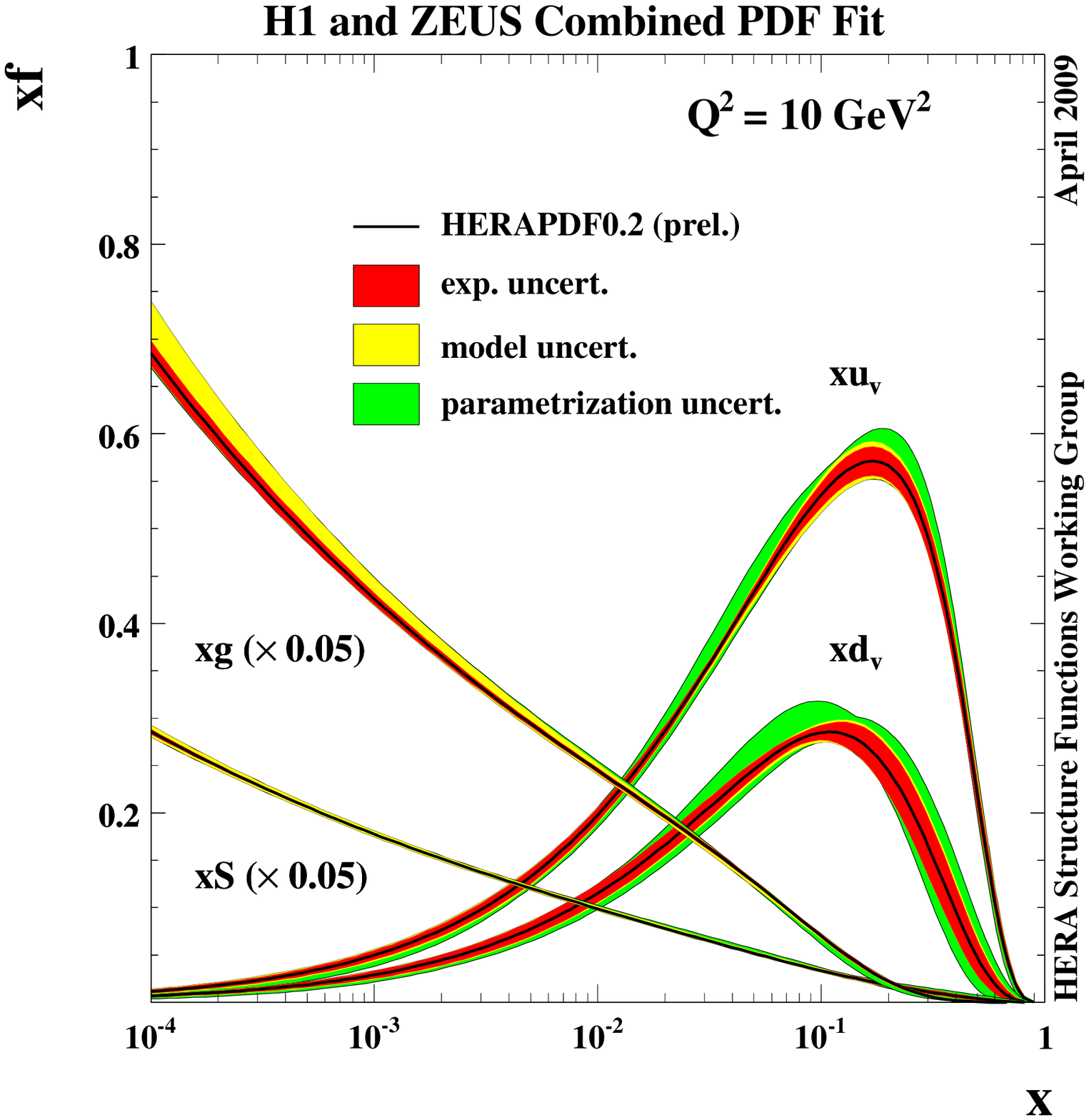,width=6.8cm}
  \vspace*{-10mm}
  \end{minipage}\\
\caption{
{\it Left} : BBS predictions for various statistical unpolarized parton distributions versus $x$ at $Q^2=10\mbox{GeV}^2$. {\it Right} : Parton distributions at $Q^2=10\mbox{GeV}^2$, as determined by the H1PDF fit, with different uncertainties (Taken from Ref. \cite{aaron}).}
\label{fi:fig1}
\end{figure}
For illustration, we will just give one recent result, directly related to the determination of the quark distributions from unpolarized DIS.
 We display on Fig.~1($\it{Left}$), the resulting unpolarized statistical PDF versus $x$ at $Q^2$=10 $\mbox{GeV}^2$, where $xu_v$ is the $u$-quark valence, $xd_v$ the $d$-quark valence, with their characteristic maximum around $x=0.3$, $xG$ the gluon and $xS$
stands for twice the total antiquark contributions, $\it i.e.$ $xS(x)=2x(\bar {u}(x)+ \bar {d}(x) + \bar {s}(x))+ \bar {c}(x))$. Note that $xG$ and $xS$ are downscaled by a factor 0.05. They can be compared with the parton distributions as determined by the H1PDF 2009 QCD NLO fit, shown also in Fig.~1($\it{Right}$), and the agreement is rather good. The results are based on recent $ep$ collider data from HERA, combined with previously published data and the accuracy is typically
in the range of 1.3 - 2$\%$.\\

\begin{figure}[htb]
  \begin{minipage}{7.0cm}
    \vspace*{+10mm}
  \epsfig{figure=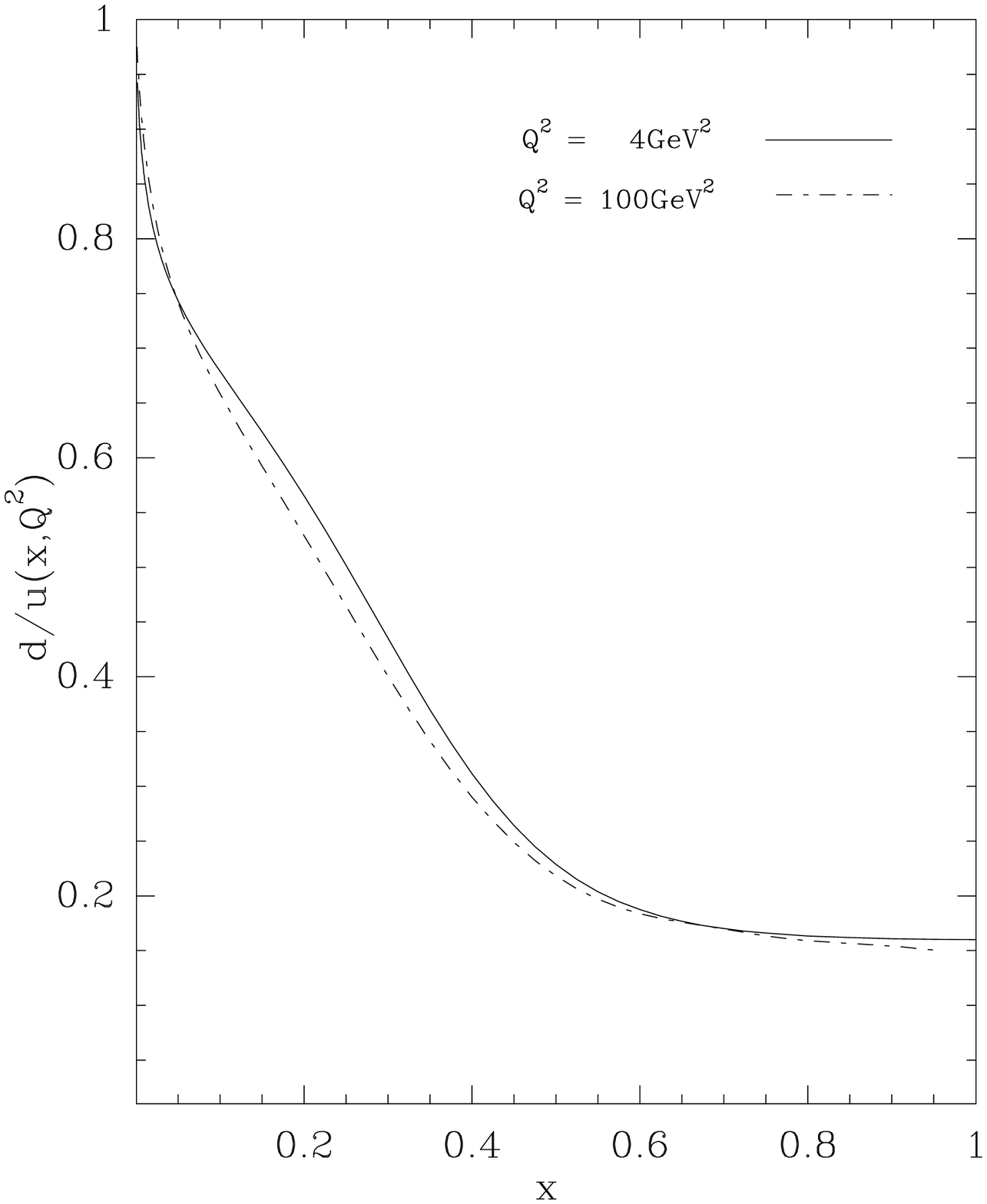,width=7.4cm}
  \end{minipage}
  \vspace*{+3mm}
    \begin{minipage}{7.0cm}
  \epsfig{figure=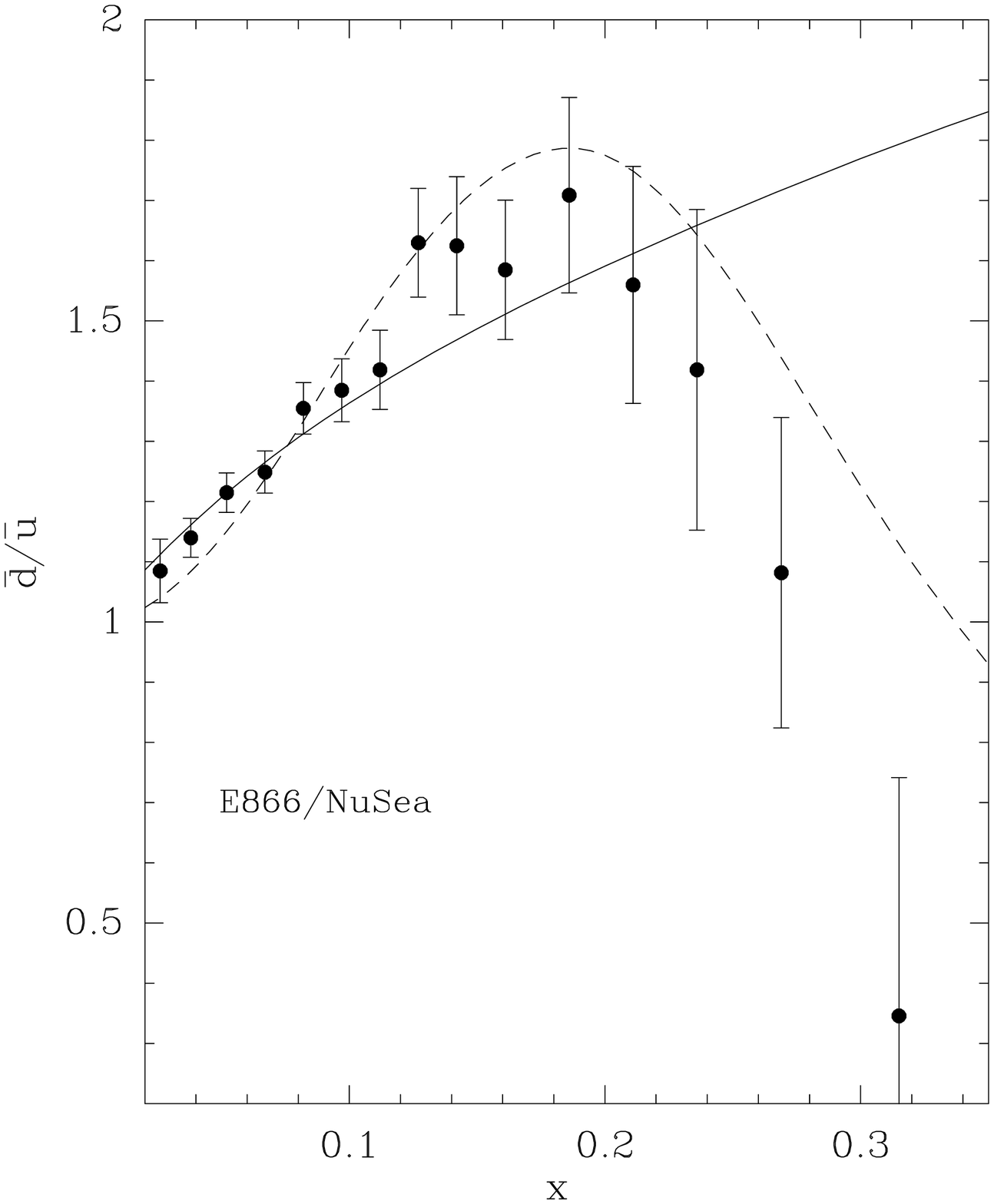,width=6.8cm}
  \vspace*{-10mm}
  \end{minipage}\\
  \caption{{\it Left} : The ratio $d(x)/u(x)$ as function of $x$ for $Q^2 = 4\mbox{GeV}^2$ 
(solid line) and $Q^2 =100\mbox{GeV}^2$ (dashed-dotted line). {\it Right} : Comparison of the data on $\bar d / \bar u (x,Q^2)$ from E866/NuSea
at $Q^2=54\mbox{GeV}^2$
\cite{E866}, with the prediction of the statistical model (solid curve) 
and the set 1 of the parametrization proposed in Ref. \cite{Sassot}
(dashed curve).}
\label{fi:doveru}
\end{figure}

 Another interesting 
point concerns the behavior of the ratio $d(x)/u(x)$, 
which depends on the mathematical properties of the ratio of two Fermi-Dirac 
factors, outside the region dominated by the diffractive contribution. 
So for $x>0.1$, this ratio is expected to decrease faster for 
$X_{0d}^+ - \bar x < x < X_{0u}^+ + \bar x$ and then above, for 
$x > 0.6$ it flattens out.
This change of slope is clearly visible in Fig.~\ref{fi:doveru} (${\it Left}$), with a very 
little $Q^2$ dependence. Note that our prediction for the large $x$ behavior,
differs from most of the current literature, namely $d(x)/u(x) \to 0$
for $x \to 1$, but we find $d(x)/u(x) \to 0.16$ near the value $1/5$,
a prediction
originally formulated in Ref.~\cite{FJ}.
This is a very challenging question, since the very high-$x$ region remains
poorly known.\\

 \begin{figure}[htb]
   \hspace*{+12mm}
  \epsfig{angle=-90,figure=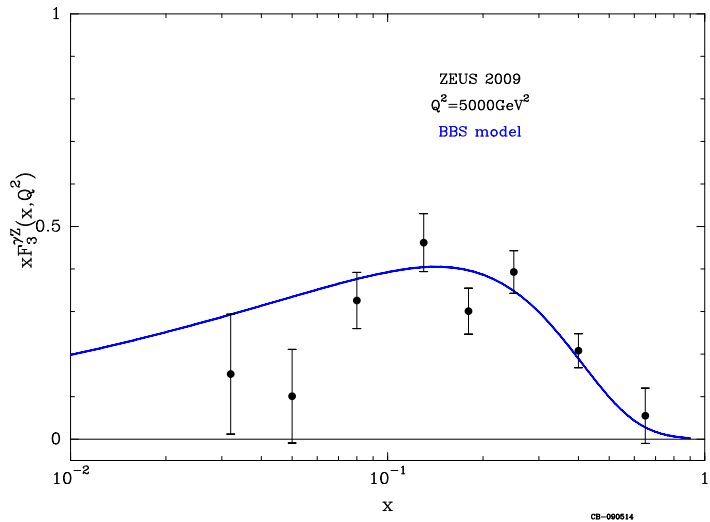,width=10.5cm}
  \vspace*{-3mm}
\caption{ The interference term $xF_3^{\gamma Z}$ extracted in $e^{\pm}p$ collisions at HERA. Data from \cite{fgz-zeus} compared to the BBS prediction..}
\label{fi:fig3}
\end{figure}
To continue our tests of the unpolarized parton distributions, we must come 
back to the important question of the flavor asymmetry of the light
antiquarks. Our determination of $\bar u(x,Q^2)$ and
$\bar d(x,Q^2)$ is perfectly consistent with the violation of the Gottfried
sum rule, for which we found the value $I_G= 0.2493$ for $Q^2=4\mbox{GeV}^2$.
Nevertheless there remains an open problem with the $x$ distribution
of the ratio $\bar d/\bar u$ for $x \geq 0.2$.
According to the Pauli principle, this ratio is expected to remain above 1 for any value of
$x$. However, the E866/NuSea Collaboration \cite{E866} has
released the final results corresponding to the analysis of their full
data set of Drell-Yan yields from an 800 GeV/c proton beam on hydrogen
and deuterium targets and they obtain the ratio, for $Q^2=54\mbox{GeV}^2$, 
$\bar d/\bar u$ shown in Fig.~\ref{fi:doveru} ({\it Right}). 
Although the errors are rather large in the high-$x$ region,
the statistical approach disagrees with the trend of the data.
Clearly by increasing the number of free parameters, it
is possible to build up a scenario which leads to the drop off of
this ratio for $x\geq 0.2$.
For example this was achieved in Ref. \cite{Sassot}, as shown 
by the dashed curve in Fig.~\ref{fi:doveru} ({\it Right}). There is no such freedom in the statistical
approach, since quark and antiquark distributions are strongly related. On the experimental side, there are now new
opportunities for extending the $\bar d/ \bar u$ measurement to larger $x$ up to $x=0.7$, 
with the upcoming E906 experiment at the 120 GeV Main Injector at Fermilab \cite{E906} and a proposed
experiment at the new 30-50 GeV proton accelerator at J-PARC \cite{JPARC}.\\
One can also test the behavior of the interference term between the photon and the $Z$ exchanges, which can be isolated in neutral current $e^{\pm}p$ collisions at high $Q^2$. We have to a good approximation, if sea quarks are ignored, $xF_3^{\gamma Z}= \frac{x}{3}(2u_v + d_v)$ and the comparison between data and prediction is displayed in Fig.~\ref{fi:fig3}. Here again, we note the remarkable agreement in the
high-$x$ region.\\
\begin{figure}[htb]
  \begin{minipage}{7.0cm}
  \epsfig{figure=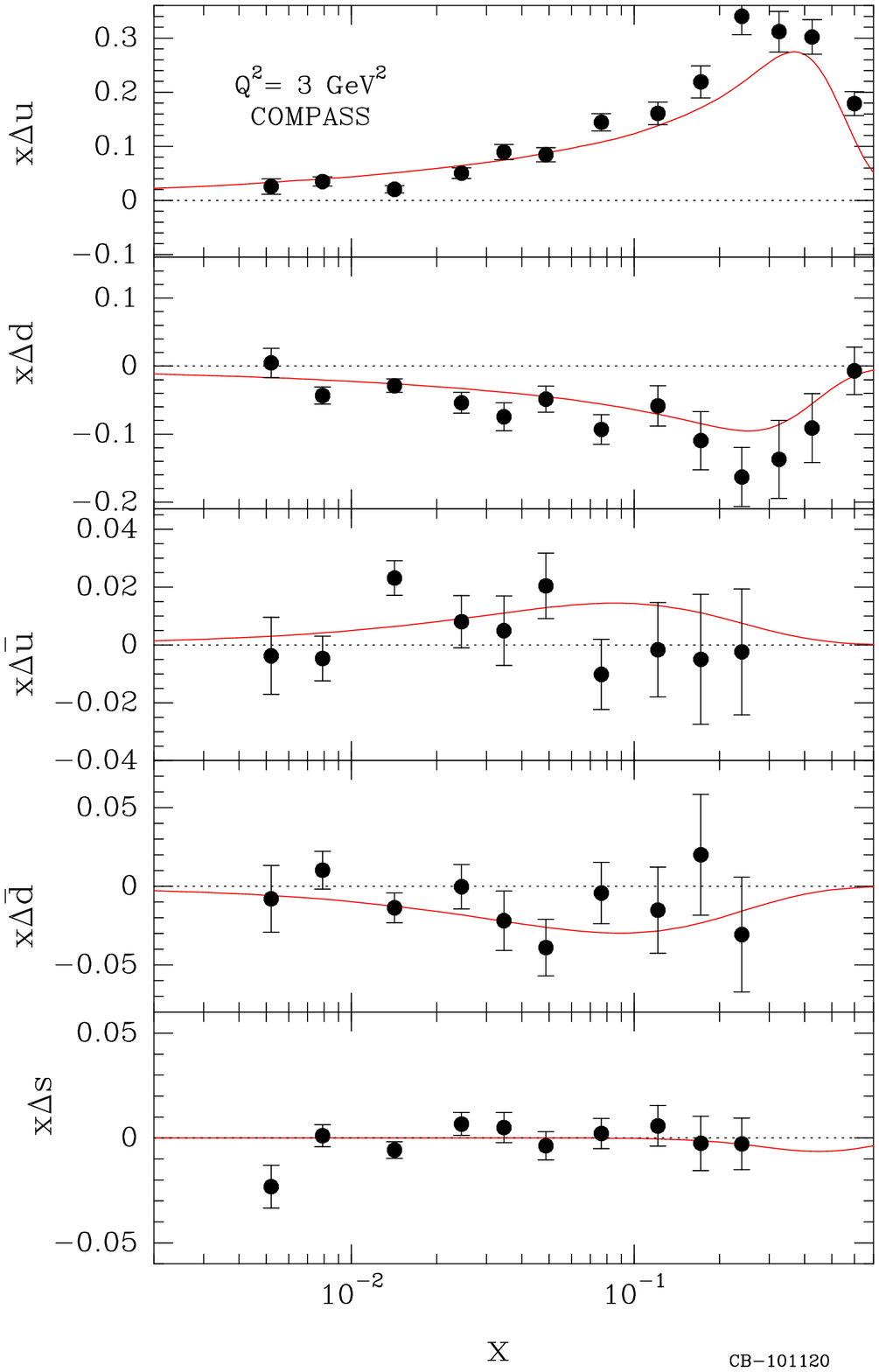,width=7.4cm}  
  \end{minipage}
    \begin{minipage}{7.0cm}
  \epsfig{figure=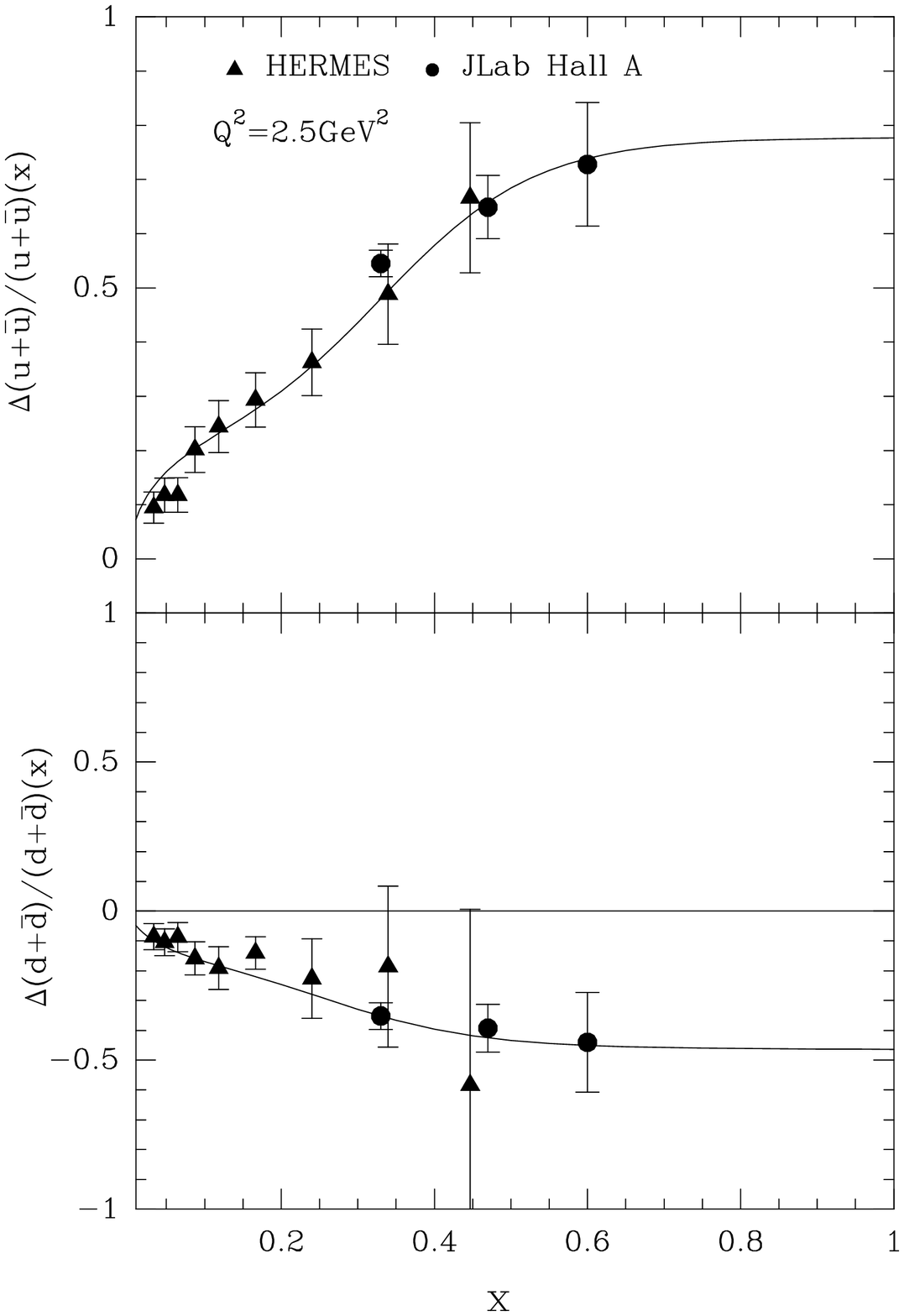,width=7.8cm}
  \vspace*{-10mm}
  \end{minipage}\\
\caption{{\it Left} : Quark and antiquark helicity distributions as a function of $x$ for $Q^2 = 3\mbox{GeV}^2$. Data from COMPASS \cite{COMPASS}. The curves are predictions from the statistical approach. {\it Right} : Ratios $(\Delta u + \Delta \bar u)/(u + \bar u)$ and 
$(\Delta d + \Delta \bar d)/(d + \bar d)$ as a function of $x$.
Data from Hermes for $Q^2 = 2.5\mbox{GeV}^2$ \cite{herm99} and
a JLab Hall A experiment \cite{JLab04}. The curves are predictions from the statistical approach.}
\label{fi:fig4}
\end{figure}

Analogous considerations can be made for the corresponding helicity 
distributions, whose most recent determinations are shown in Fig.~\ref{fi:fig4} ($\it {Left}$).
By using a similar argument as above, the ratio $\Delta u(x)/u(x)$ 
is predicted to have a rather fast increase in the $x$ range 
$(X^-_{0u}-\bar{x},X^+_{0u}+\bar{x})$
and a smoother behaviour above, while $\Delta d(x)/d(x)$, which is negative,
has a fast decrease in the $x$ range $(X^+_{0d}-\bar{x},X^-_{0d}+\bar{x})$ 
and a smooth one above. This is exactly the trends displayed in 
Fig.~\ref{fi:fig4} ($\it Right$) and our predictions are in perfect agreement
with the accurate high-$x$ data. We note the behavior near $x=1$, another typical property of the statistical
approach, is also at variance with predictions of the current literature. 
The fact that $\Delta u(x)$ is more concentrated in the higher $x$ region than
$\Delta d(x)$, accounts for the change of sign of $g^n_1(x)$, which becomes
positive for $x>0.5$, as first observed at Jefferson Lab \cite{JLab04}.\\

\begin{figure}[htb]
  \hspace*{+12mm}
  \epsfig{angle=-90,figure=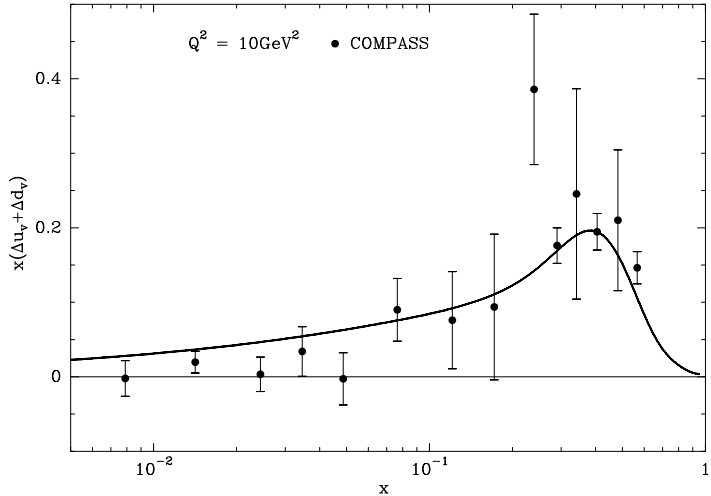,width=10.5cm}
  \vspace*{-3mm}
\caption{ The valence quark helicity distributions versus $x$ and evolved at $Q^2=10\mbox{GeV}^2$. The solid curve is the BBS prediction
of the statistical approach and the data come from Ref. \cite{compass}.}
\label{fi:fig5}
\end{figure}

Concerning the light antiquark helicity distributions, the statistical 
approach
imposes a strong relationship to the corresponding quark helicity
distributions. In particular, it predicts $\Delta \bar u(x)>0$ and $\Delta \bar
d(x)<0$, with almost the same magnitude, in contrast with the
simplifying assumption $\Delta \bar u(x)=\Delta \bar d(x)$, often adopted in
the literature. The COMPASS experiment
at CERN has measured the valence quark helicity distributions, defined as
$\Delta q_v(x)= \Delta q(x)-\Delta \bar q(x)$. These recent results displayed
in Fig.~\ref{fi:fig5} are compared to our prediction and the agreement is best in the
high-$x$ region.\\
The data give $\Delta \bar u(x) + \Delta \bar d(x) \simeq 0$, which implies either small or
opposite values for $\Delta \bar u(x)$ and $\Delta \bar d(x)$. Indeed $\Delta
\bar u(x)>0$ and $\Delta \bar d(x)<0$, predicted by
the statistical approach \cite{bbs1} (see Fig.~\ref{fi:fig4} ({\it Left}), lead to a non negligible
 positive contribution of the sea to the Bjorken sum rule, an interesting consequence.

 We now turn to another important aspect of the statistical PDF and very briefly discuss
a new version of the extension to the transverse momentum dependence (TMD).
In Eqs.~(\ref{eq1},\ref{eq2}) the multiplicative factors $X^{h}_{0q}$ and
$(X^{-h}_{0q})^{-1}$ in
the numerators of the non-diffractive parts of $q$'s and $\bar{q}$'s
distributions, imply a modification
of the quantum statistical form, we were led to propose in order to agree with
experimental data. The presence of these multiplicative factors was justified
in our earlier attempt to generate the TMD \cite{bbs5}, but it was not properly done and 
a considerable improvement was achieved recently \cite{bbs6}. We have introduced some thermodynamical
potentials $Y^h_{0q}$, associated to the quark transverse momentum $k_T$, and
related to $X^{h}_{0q}$ by the simple relation
$\mbox{ln}(1+\exp[Y^h_{0q}])=kX^h_{0q}$. We were led to choose k=3.05 and this method involves another parameter
$\mu^2$, which plays the role of the temperature for the transverse degrees of
freedom and whose value was determined by the transverse energy sum rule.
 We have calculated the $p_T$ dependence of semiinclusive DIS double
longitudinal-spin asymmetries, taking into account the effects of the
Melosh-Wigner rotation, for $\pi^{\pm}$ production, which were compared
to recent experimental data from CLAS at JLab.\\

A new set of PDF was constructed in the framework of a statistical
approach of the nucleon.
All unpolarized and polarized distributions depend upon {\it nine}
free parameters for light quarks and gluon, with some physical meaning.
 New tests against experimental (unpolarized and polarized)
data on DIS, Semi-inclusive DIS and also hadronic processes, are very satisfactory.
It has a good predictive power, but some special features remain to be verified, specially in the high-$x$ region, a serious challenge
for the future.\\

{\bf Acknowledgments}\\
I am grateful to the organizers of HiX2010 for their warm hospitality at JLab and for their invitation to present this talk. My special thanks go to Dr. Simona Malace for providing a partial financial support and for making, this meeting so successful.


\newpage
\begin{thebibliography}{99}

\bibitem{bbs1} C. Bourrely, F. Buccella and J. Soffer,
\emph{Eur. Phys. J. C} {\bf 23}, 487 (2002).

\bibitem{bbs2} C. Bourrely, F. Buccella and J. Soffer,
\emph{Phys. Lett. B} {\bf 648}, 39 (2007).

\bibitem{bbs3}  C. Bourrely, F. Buccella and J. Soffer,
\emph{Mod. Phys. Lett. A} {\bf 18}, 771 (2003).

\bibitem{bbs4} C. Bourrely, F. Buccella and J. Soffer,
\emph{Eur. Phys. J. C} {\bf 41}, 327 (2005).

\bibitem{aaron} F. D. Aaron {\it et al.}, [H1 Collaboration], 
\emph{Eur. Phys. J. C} {\bf 64}, 561 (2009).

\bibitem{FJ} G.R. Farrar and D.R. Jackson, \emph{Phys. Rev. Lett.} {\bf 35}, 1416 
(1975).

\bibitem{E866} R.S. Towell {\it et al.}, [FNAL E866/Nusea Collaboration], 
\emph{Phys. Rev. D} {\bf 64}, 052002 (2001).

\bibitem{Sassot} A. Daleo, C.A. Garc\'\i a Canal, G.A. Navarro and R. Sassot,
\emph{Int. J. Mod. Phys. A} {\bf 17}, 269 (2002).

\bibitem{E906} E906 Collaboration, D.F. Geesaman {\it et al.}, FNAL Proposal E906, April 1, 2001.

\bibitem{JPARC} J.C.. Peng {\it et al.}, hep-ph/0007341.

\bibitem{fgz-zeus}
S. Chekanov, {\it et al.}, [ZEUS Collaboration], \emph{Euro. Phys. J. C} {\bf 62}, 625 (2009).

\bibitem{COMPASS} M. Alekseev {\it et al.} [COMPASS Collaboration], \emph{Phys. Lett. B} {\bf 693}, 227 (2010).

\bibitem {herm99} K. Ackerstaff {\it et al.}, [Hermes Collaboration], 
\emph{Phys. Lett. B} {\bf 464}, 123 (1999).


\bibitem{JLab04} X. Zheng {\it et al.}, [Jefferson Lab Hall A Collaboration], 
\emph{Phys. Rev. C} {\bf 70}, 065207 (2004).

\bibitem{compass} M. Alekseev {\it et al.} [COMPASS Collaboration], \emph{Phys. Lett. B} {\bf 660}, 458 (2008).

\bibitem{bbs5} C. Bourrely, F. Buccella and J. Soffer,
\emph{Mod. Phys. Lett. A} {\bf 21}, 143 (2006).

\bibitem{bbs6} C. Bourrely, F. Buccella and J. Soffer, arXiv:1008.5322v1 [hep-ph].

\end{thebibliography}
\end{document}